\documentclass[proceedings]{JHEP} % 10pt is ignored!

\usepackage{epsfig}         % please use epsfig.
\newbox\mybox
\newcommand\fverb{\setbox\mybox=\hbox\bgroup\verb}
\newcommand\fverbdo{\egroup\medskip\noindent\fbox{\unhbox\mybox}\ }
\newcommand\fverbit{\egroup\item[\fbox{\unhbox\mybox}]}
\newcommand{\va}[1]{\langle{#1}\rangle}
\newcommand{\gev}[1]{\relax\ifmmode{\mbox{GeV}^{#1}}\else{GeV$^{#1}${ }}\fi}
\def\Gev{\relax\ifmmode{\mbox{GeV}}\else{GeV{ }}\fi}
\def\Mev{\relax\ifmmode{\mbox{MeV}}\else{MeV{ }}\fi}

\font\beeg=cmr17 scaled 1600        % Stylish initials
\newcommand\init[1]{\setbox\mybox=\hbox{{\beeg #1}~}%
           \noindent\global\hangindent=\wd\mybox\global\hangafter-2%
           \sc\smash{\llap {\lower 13.2pt \box\mybox}}}
\title{New shapes of the $\rho$-meson light-cone distribution amplitudes:
       how can they influence the $B \to \rho e \nu$ decay form factors}

\author{A.\,Bakulev, S.\,Mikhailov, and R.\,Ruskov\\
        Bogoliubov Lab. of Theoretical Physics, JINR, 141980 Dubna, Russia\\
    Email: \email{bakulev@thsun1.jinr.ru}, \email{mikhs@thsun1.jinr.ru},
\email{ruskovr@thsun1.jinr.ru}}

\conference{Heavy Quark Physics 5, Dubna, Russia, 6-8 April 2000}

\abstract{We present new models of the $\rho$-meson leading-twist
light-cone distribution amplitudes based on the QCD sum rule approach
with nonlocal condensates.
Their shapes differ noticeably from that known in the literature.
The phenomenological consequences for physically important process
$B \to \rho e \nu$ are discussed in the framework
of the light-cone sum rules.
The results are compared with those found recently
by P.Ball and V.M.Braun (1997).}

\begin{document}

\section{Introduction}
The physics of $B$-decay is an attractive field
both from theoretical and experimental point of view.
Among other important problems, the extraction
of the CKM matrix elements from experimental data
has received much attention as these elements
determine our fundamental knowledge of the Standard Model.
In this context the semileptonic $B$-decays to light hadrons
($\pi$, $\rho$) were mentioned as a suitable tool
to measure the $|V_{ub}|$ (cf., e.g., \cite{BBr97}).
Recently, the CLEO collaboration
has confirmed the first experimental measurements \cite{CLEO}
of the branching ratio for $B \to \rho l \nu$
and has presented first results for the $t$-dependence
of the form factors.

As usual in QCD, one may hope that
for a quark mass heavy enough and/or large momentum tran\-sfer,
the large scale introduced would determine a perturbative regime
that presumably would simplify the physical picture.
 However, due to quark confinement
and especially for a heavy-to-light transitions (like $b \to u$),
the analysis inevitably involves the (nonperturbative) dynamics
of the light degrees of freedom.
Thus, to disentangle the properties of the heavy
quark, one is forced carefully to separate perturbative
and nonperturbative effects.

The method of QCD Sum Rules (SRs) seems to be well suited
for such separation \cite{SVZ}.
The effects of nonperturbative long-distance dynamics are accumulated
into {\it universal} objects like vacuum condensates and, more generally,
in hadronic distribution amplitudes (DAs) and bilocal correlators
(see, $e.g.$ \cite{BaBrKolesnichenko,RR96,BBr97}).
In principle, the standard QCD SR approach
implies investigation of a suitable 3-point correlator \cite{NeRad},
and usually the first few terms of the operator product expansion (OPE)
are involved. For the case of $B \to \rho$ weak transitions such a
program was performed in Refs.\cite{Ball,AliBrSim}.

However, the kinematic region of interest for the momentum transfer
to the lepton pair is quite large: $0 < t < 20.3~\gev{2}$,
and one can encounter specific problems of the approach.
Indeed, in the region of maximum recoil
to the final light meson (this corresponds to momentum transfers
$t\approx 0$, $i.e.$ far from the threshold $t^{th} \sim m_b^2$)
the relevant OPE
for the 3-point correlator becomes poorly convergent,
the correction terms being proportional to the positive
(and growing with dimension of the condensate)
powers of the heavy quark mass~\cite{AliBrSim,BBr97}.
This situation is in full correspondence with the previously investigated
transition form factor $\gamma^*(Q)\gamma^*(q) \to \pi^0$ in the
kinematics $Q^2 \gg q^2 \ge 1~\gev{2}$~\cite{MiRad89,RR96}.

In this case one has to sum up the OPE series
which naturally amounts to the nonlocal condensates (NLC)
\cite{MiRad89,MiRad92}.
These objects enter into different SRs
and may be extracted from the relevant analysis.

On the other hand, in the above-mentioned kinematics
the light-cone region dominates,
which corresponds to the heavy quark perturbative propagation.
A possible remedy of the problems with 3-point SRs
was offered within the Light Cone (LC) SR
approach \cite{BaBrKolesnichenko,CZ90,AliBrSim,BBr97}.
In this case one deals with an amplitude
in which the final hadron is already represented by its
DAs of leading twists.
In comparison to the previous approach
this amounts to an effective summation of the above-mentioned OPE series
with the price of introducing other nonperturbative quantities --
the DAs of the light hadrons.
These DAs are {\it universal} quantities, they enter
as important ingredients into the ``factorization" formalism
\cite{EfrRad80} for any hard exclusive reactions
involving hadrons.

In the remaining of this short talk we shall present new results concerning
the leading twist 2 DAs for a longitudinally and transversely polarized
$\rho$-meson obtained from QCD SRs with NLCs.
We estimate the influence of the new shapes of the DAs for
the phenomenologically important weak form factors of the $B \to \rho$
transition using the LC SR in the leading twist
approximation~\cite{AliBrSim,BBr97}.

\section{The leading twist DAs of $\rho$-meson}

Here, we discuss the light-cone DAs of the leading twist
for the $\rho$-meson.
At least for the leading twist DAs,
a physical quark-parton interpretation exists:
it is a nonperturbative amplitude for a hadron to decay into
collinear quark(s)-anti\-quark(gluon).

The DAs under consideration,
$\varphi^L_{\rho}(x)$, $\varphi^T_{\rho}(x)$,
parameterize the gauge-invariant matrix elements
with the $\rho(770)$-meson ($J^{PC} =1^{--}$)
of the (nonlocal) vector current ($\mu^2$ is the factorization scale),
\begin{eqnarray}
\lefteqn{\va{0\mid\bar u(z)E(z,0)\gamma_{\mu}d(0)\mid \rho_{L}(p)}
\Big|_{z^2=0}=  \label{eq-Rho_L}}\nonumber\\
&& \qquad\qquad = if_{\rho}^{L}p_{\mu}
   \int^1_0 dx\ e^{ix(zp)}\ \varphi^L_{\rho}(x,\mu^2)\nonumber,
% \label{eq-Rho_L}
\end{eqnarray}
and\footnote{For $p_z \to \infty$ we incorporated that
$\varepsilon_{\mu}^{\lambda=0}\simeq i p_{\mu}/m_{\rho}$.} the
tensor current
\begin{eqnarray}
\lefteqn{\va{0\mid\bar u(z)E(z,0)\sigma_{\mu\nu}d(0)\mid \rho_{\perp}(p)}
\Big|_{z^2=0}=\label{eq-Rho_T}}\nonumber\\
&& = if_{\rho}^{T}
  \left(\varepsilon_{\mu}^{\perp}p_{\nu}
       -\varepsilon_{\nu}^{\perp}p_{\mu}\right)
   \int^1_0 dx\ e^{ix(zp)}\ \varphi^T_{\rho}(x, \mu^2)\nonumber ,
% \label{eq-Rho_T}
\end{eqnarray}
The first estimates of the nontrivial moments,
$\va{\xi^N} \equiv\int_0^1\varphi(x)(2x-1)^N\,dx$,
of these functions were obtained by
Chernyak\&Zhitnitsky (CZ)~\cite{CZ84} using the QCD SR
for suitable 2-point current correlators
of the vector (tensor) currents with the derivatives.
A detailed revision of these results within the standard approach
were presented by Ball\&Braun (BB)\cite{BBr96}.
The analysis was also extended by introducing the DAs
of nonleading twist (3, 4) and incorporating equations of motion
(see, e.g., \cite{AliBrSim,RR96}).
In recent papers \cite{BBrKT,BBr98}, this so-called standard analysis
was completed by taking into account the finite mass corrections
as well. Note, that in the framework of the approach
one should restrict oneself
to an estimate of the 2-nd moment $\va{\xi^2}$ of the DA
to restore its shape\footnote{%
We should remark in this respect that the standard approach could not
provide a reliable estimate even for the 2-nd moment of DA,
see \protect{\cite{MiRad92,BM98,Rad98}}}.

 We would like to emphasize that
the standard QCD SR approach for the nontrivial moments of the DAs
encounters similar problems as mentioned above in the case of 3-point SRs.
The relevant OPE for the $N$-th moment receives
an $N$ power enhancement, and the higher
is the dimension of the operators involved in the OPE
the stronger power growth is observed.
Thus, the OPE for higher moments is poorly convergent and the
evaluation of the moments hardly make sense
(see the criticism in \cite{MiRad89,RR96,BM98}).

It was recognized~\cite{MiRad92} that such an $N$ enhancement
is a consequence of expanding the originally NLCs,
like $\va{\bar{q}(0)E(0,z)q(z)}$,
into the local ones
$\va{\bar{q}(0) q(0)}$, $\va{\bar{q}(0)\nabla^2 q(0)}$, etc.,
appearing in OPE.
Physically this means that the correlation length
of the vacuum fluctuations, $\lambda_q^{-1}$,
was supposed to be much larger than the typical hadronic scale
$\sim m_{\rho}^{-1}$ which appears to be an unrealistic approximation.
On the contrary, keeping the NLCs unexpanded
one would obtain a decreasing $N$ dependence
for the condensate contributions
just as it is the case with the leading perturbative term
of the OPE~\cite{MiRad92}.
On the basis of a simple model for NLC the authors
of Ref.~\cite{MiRad92} reanalyzed
the moment's QCD SR for the pion leading twist DA
and obtained a form which is rather close
to the asymptotic one $\varphi^{as}(x)=6x(1-x)$.
This result was in contrast to the double-humped
form originally suggested by CZ.
The closeness of the pion DA to its asymptotic form
at a low normalization point was supported later
using different theoretical approaches
\cite{RR96,BJ97,PPRGW}
and also from the experiment~\cite{CLEOggpi}.
Here we present results for the two leading twist DAs
of the $\rho$-meson  using the same method and essentially
the same models for the nonlocal condensates involved.
Instead of going to details, we just briefly mention
some essential features of the (NLC) QCD SRs
for the relevant quantities.

The first five moments  $N=2,4,6,8,10$
have been obtained with a good
accuracy for the DA of the longitudinally polarized
$\rho$-meson\cite{BM98}.
The shape of the DA, $\varphi_\rho^{L}(x,\mu^2)$,
restored with these moments (at $\mu^2\simeq 1~\gev{2}$),
is well established:
\begin{eqnarray}
\lefteqn{\varphi_\rho^{L}(x)=
\varphi^{as}(x)\times
\label{eq:mod_lro}} \\
&& \qquad\times\!\left(1+0.077\,C^{3/2}_2(\xi)
        -0.077\,C^{3/2}_4(\xi)\right),\nonumber
%\label{eq:mod_lro}
\end{eqnarray}
where $\xi\equiv 1-2x$, and $C^{\nu}_{n}(\xi)$ are the
Gegenbauer polynomials.
It does not differ strongly from that,
obtained in the standard way \cite{BBr96},
on the basis of a crude estimate of the second moment only.
Nevertheless, one may observe an essential difference
in the end-point behavior, numerically revealing itself
in the important inverse moment of DA: $\int_{0}^{1}
\frac{\varphi_\rho^{L}(x)}{x}\ dx = 3 (\mbox{here})$,
$3.54 (\mbox{B\&B})$, $4.38 (\mbox{C\&Z})$.

The case of transversally polarized $\rho$-meson is more peculiar
because the tensor current is of mixed P-parity
and projects also on states with $J^{PC} =1^{+-}$
(the lowest resonance being the $b_1(1235)$-meson).
The relevant correlator of two tensor currents
$\Pi^{\mu\nu;\alpha\beta}_N(q)$
contains different invariant form factors
at the corresponding independent tensor structures,
which, in general, can contaminate contributions
from both the types of hadronic states.

In fact, in Refs.\cite{CZ84,BBr96}, a mixed-parity SR was investigated
based on the projection over $z^{\nu}z^{\beta}g^{\mu\alpha}$.
The feature of this SR is that the contribution
of the four-quark condensate is absent.
On the other hand,  this SR
receives a numerically strong contribution
from the gluon condensate and, in fact, it should be
sensitive to the model of the nonlocal entity, that is
still ill-known, contrary to the quark case.

Thus, it is suggested to use, instead,
a pure parity SR which relates only
to states of definite parity ($\rho$, $\rho'$, etc. as $P=-1$).
Such NLC SR allows one to extract not only tensor coupling
$f^T_{\rho}$ \cite{BBr96}, but
also the higher moments for $\va{\xi^N}^T_{\rho}$
with $N=2,4,6,8$. It should be noted that contrary to the
longitudinal case, the higher moments are far from their asymptotic values.
The model for the DA $\varphi^T_{\rho}(x,\mu^2\simeq 1~\gev{2})$
reads:
\begin{eqnarray}
\lefteqn{\varphi_\rho^{T}(x)=
 \varphi^{as}(x)\!\times\!\left(1+0.29\,C^{3/2}_2(\xi)+{}\right.
\label{eq:mod_tro}} \\
&&\left.\qquad\qquad\quad{}+0.41\,C^{3/2}_4(\xi)-0.32\,C^{3/2}_6(\xi)\right) ,
\nonumber
%\label{eq:mod_tro}
\end{eqnarray}
In Fig.1, we have plotted our DA
$\varphi_\rho^{T}(x)$ in comparison
with that proposed by Ball\&Braun \cite{BBr96}.
One may observe an essential difference in the shape
and especially for the end-point behavior.

The  oscillatory form of our model DA
is not an artifact of a by hand truncation of the series
in Gegenbauer polynomials.
In fact, using the higher moments obtained,
we are able to calculate also the nonperturbative coefficients
of the higher polynomial(s), which occurred to be very small.
It is worth mentioning that the better
\FIGURE{\epsfig{file=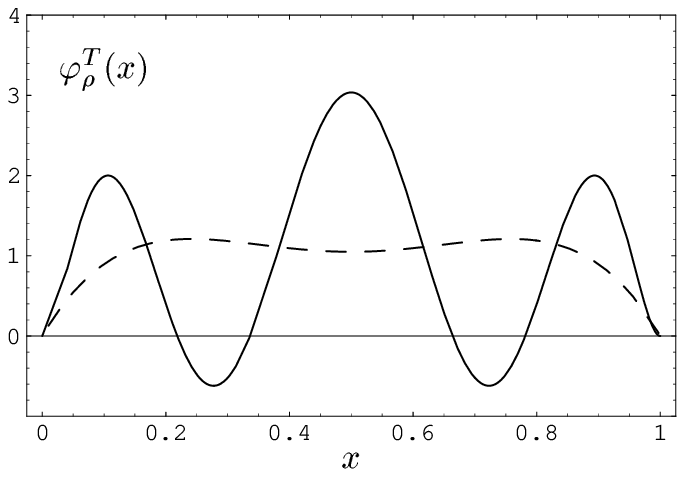,width=5cm}%
\caption{$\varphi_{\rho}^{T,mod}(x,1~\gev{2})$:
our model -- solid line, B\&B model -- dashed line.}\label{wf-rot}}
knowledge of the end-point
region in our NLC SR approach is a consequence of the ability to extract
the higher moments with enough good accuracy.
Such a feature was, actually, expected because with the
NLCs at hand our knowledge on the OPE side of the SR
increases. Correspondingly, one also may be able to extract a more detailed
information on the phenomenological side. In fact, the mass $m_{\rho'}$,
the decay constant(s) $f_{\rho'}^L$, $f_{\rho'}^T$,
and the first 4 moments for the radially excited $\rho^{'L}$ was obtained
(cf. \cite{BM98} for details).

\section{The $B \to \rho e \nu$ transition form factors within the
light-cone SR}

The relevant invariant
form factors of the independent Lorentz structures corresponding to
the transition matrix element $\va{\rho,\lambda|(V-A)_{\mu}|B}$ are denoted
as $V(t)$, $A_1(t)$, and $A_2(t)$.
As mentioned in the Introduction, the LC SR were proved to be a suitable
approach to the semileptonic transition form factors, especially for the
region of maximum recoil~\cite{BBr97}.
The ``theoretical'' side of the
LC SR can be expressed as a convolution of a short distance coefficient
function $CF(m_b,t,p_B^2;x,\mu^2)$
corresponding to the propagation of the heavy quark and the
leading twist $\rho$-meson DAs.

In principle, it may receive corrections,
both perturbative and nonperturbative.
The $\alpha_s$-cor\-rections
to the hard part $CF(\ldots)$ as well as the contributions
of higher twist (3 and 4) 2- and 3-body DAs amplitudes
were investigated in detail in Refs.~\cite{BBrKT,BBr98}.
The latter also include the ``kinematic'' higher twist corrections
due to finite $\rho$-meson mass~\cite{BBr98}.
Not going to a detailed discussion of these comprehensive works,
we mention that as a net result the impact of the $\alpha_s$-corrections
is on the level of $5\%$ for relatively small momentum transfers,
and the contribution of the higher twists is at most
$3\%$ (cf. \cite{BBr98}).

Thus, to estimate the influence of the new nonperturbative
input presented in the previous section, we have used the LC SR in the
leading twist approximation (cf. \cite{BBr97}).
Just as in the case of the LC expansion
for the transition amplitude $\gamma^*\gamma \to \pi^0$,
one might expect high sensitivity to the end-point behavior
of the DAs as they enter into convolution integrals like
$\int_0^1 dx \varphi(x)/x$.

However, there are some essential differences which effectively soften
our expectations. First, the DAs also enter into the
``phenomenological'' side of the SR in the ``continuum'' contribution of
higher excited states in the channel with $B$-meson quantum numbers.
This, actually, is a specific feature of any LC SR. By subtracting the
``continuum'' one actually obtains ``infrared safe quantities'' like
$\int_{\epsilon}^1 dx \varphi(x)/x$ where
$\epsilon\simeq (m_b^2-t)/(s_0^B-t)$,
$m_b\simeq 4.8\,\Gev$, and $s_0^B\simeq 34~\gev{2}$ is the continuum
threshold in the $B$-channel\footnote{As we shall see
below, the LC SRs ``prefer'' a higher value.}
as defined from the 2-point QCD SRs for the
$B$-meson decay constant $f_B$ (see~\cite{Eletsky,BBBDosch}).

For $t\approx 0$, $\epsilon\simeq 0.5-0.6$ and the LC SR should not be
so sensitive to the end-point region $x \sim 0$. Obviously, the end-point
region becomes to be important for higher momentum transfers $t$.
However, for $t\ge 20~\gev{2}$ the LC expansion would hardly make sense.

The second factor which eventually decreases the importance of the end-point
region is connected with the standard Borel transformation of the SR
with respect to the virtuality of the $B$-meson current: $-p_B^2 \to M_B^2$.
The corresponding hard part $CF(\ldots)$ then produces
a standard suppression exponent: $\exp(\bar{x}(t-m_b^2)/x\,M_b^2)$.
Numerically, it occurred to be less important.

We have treated the LC SRs using the same input parameters and the same
procedure  of extracting the physical form factors as in
Ref.\cite{BBr97}. However, if one tries to fix the onset of the "continuum"
by hand to the value $s_0^B \simeq 34~\gev{2}$ dictated by the 2-point
SRs for $f_B$,
one encounters {\it inadmissible} uncertainties
in the determination of the form factors when using our new
nonperturbative input DAs. To get a stable SR, one is
forced to allow a higher value for the $s_0^B$ parameter.

All evaluated form factors  are a little bit larger
than the corresponding estimations with the B\&B leading twist DAs.
The difference becomes more pronounced for large value of the momentum
transfer $t$,
($m_b^2-t \sim {\cal O}(m_b)$). The last is not surprising due to higher
sensitivity to the end-point behavior of the input DA in this region.
The form factors presented are determined with few times better processing
accuracy with new ``optimal'' thresholds $s_0^B$.
Note that the parameters of the usual ``pole'' parameterization
of the form factors change significantly as compared to that in
\cite{BBr97}, {\it e.g.},
\begin{eqnarray}
% \label{A-1}
 A_1(t)= \frac{0.283}{1-0.157(t/m_B^2)-0.837(t/m_B^2)^2} \nonumber
\end{eqnarray}
The important form factor $A_1(t)$ (solid line)
increases about $5 - 10 \% $ in comparison with the B\&B result
(the bars in the figure show the B\&B errors),
with an optimal threshold $s_0^B \simeq 45$ GeV$^2$.
From a physical point of view one should
\FIGURE{\epsfig{file=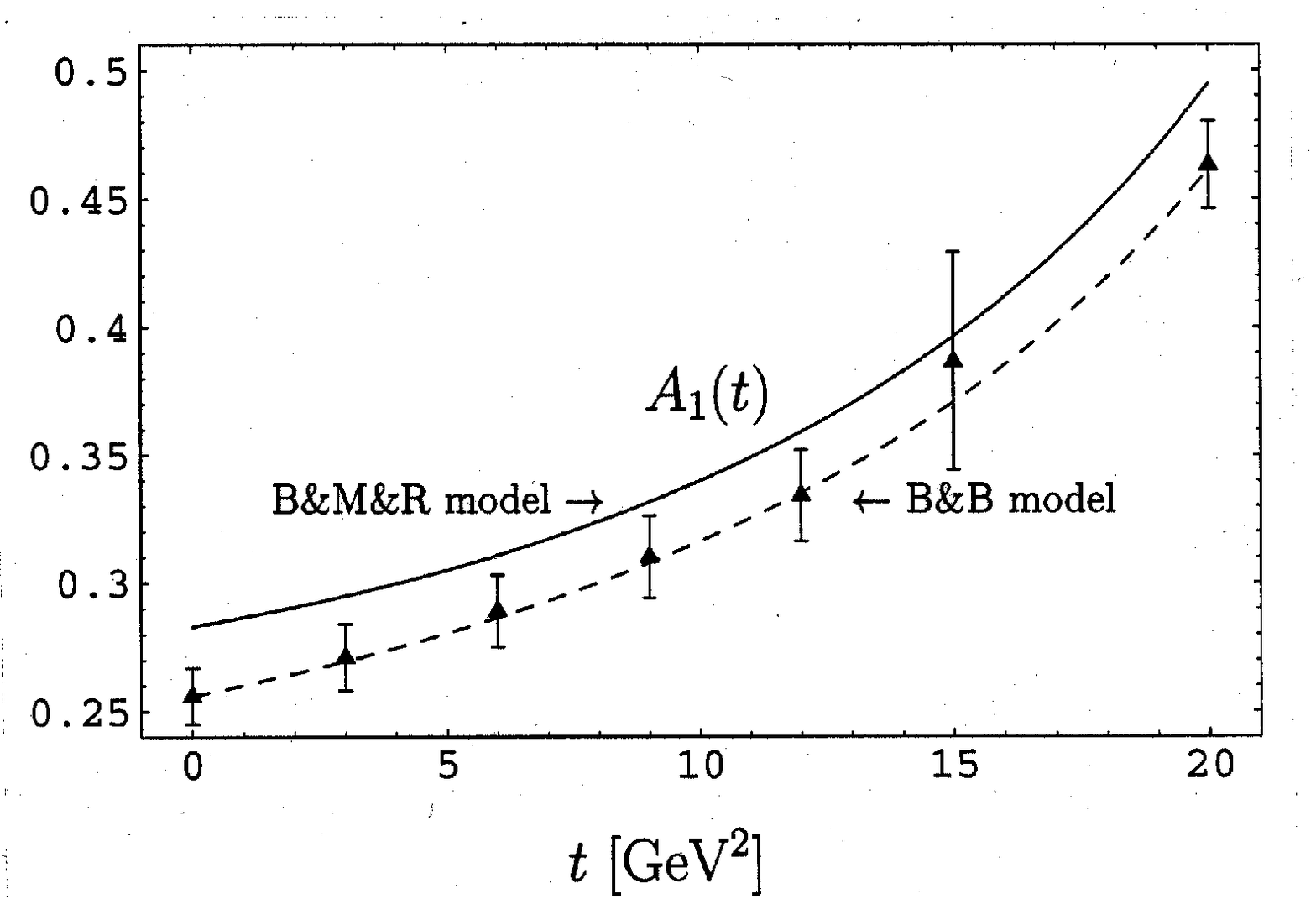,width=5cm}%
 \label{ff-a1}}
%\section{Discussion}
consider the duality interval
$s_0^B$ as a characteristic of the spectra in the $B$-channel.
Thus, in a self-consistent approach it is desirable to obtain the same
(physical) value for $s_0^B$ from different SRs.

However, the experimental information for higher excited
states in the $B$-channel is poor \cite{PDG}.
From theoretical side, the value of $f_B$ as well as $s_0^B$ was a
point of controversial issues (cf.~\cite{Eletsky,BBBDosch}).
In the most detailed analysis of the 2-point SRs for $f_B$,
the $\alpha_s$-corrections to the leading term in the OPE were
proven to be of importance~\cite{Eletsky,BBBDosch}. Actually,
in Ref.~\cite{BBBDosch}, it was argued that an effective summation
of the leading logs dictates the argument of $\alpha_s$ to be
$\sim 1~\mbox{GeV}$ rather than $\sim m_b$. As a result, the
values of $s_0^B$ from the interval $34 - 38~\gev{2}$ were preferred.
In this context, the increase of the effective threshold
$s_0^B$, as determined from the LC SR, demonstrates a deficiency
of the Light Cone SRs for the $B \to \rho$ transition
(at least, to the leading twist order).

We are grateful to V.M.Braun, H.G.Dosch, K.Goeke, M.Polyakov,
N.G.Stefanis, and C.Weiss for useful discussions.
This work was supported in part by the RFFI grant N 00-02-16696 and
by the Hei\-senberg--Landau Program.


\begin{thebibliography}{999}
%
\bibitem{BBr97}P.Ball and V.M.Braun., \prd{55}{1997}{5561}.
%
\bibitem{CLEO}CLEO Collab., J.P.Alexander et al.,
 \prl{77}{1996}{5000}; \prd{61}{2000}{5000}.
%
\bibitem{SVZ}M.A.Shifman, A.I.Vainstein and V.I.Zakharov,
\npb{147}{1979}{385}.
%
\bibitem{BaBrKolesnichenko}I.I.Balitsky, V.M.Braun
and A.V.Koles\-ni\-chen\-ko, \sjnp{44}{1986}{1028}; \npb{312}{1989}{509}.
%
\bibitem{RR96}A.V.Radyushkin and R.Ruskov, \npb{481}{1996}{625}.
%
\bibitem{NeRad}V.A.Nesterenko and A.V.Radyushkin, \plb{115}{1982}{410}.
%
\bibitem{Ball}P.Ball, \prd{48}{1993}{3190}.
%
\bibitem{AliBrSim} A.Ali, V.M.Braun and H.Simma, \zpc{63}{1994}{437}.
%
\bibitem{MiRad89}S.V.Mikhailov and A.V.Radyushkin, \sjnp{52}{1990}{697}.
%
\bibitem{MiRad92}S.V.Mikhailov and A.V.Radyushkin,
\jetpl{43}{1986}{712};
\prd{45}{1992}{1754}.
%
\bibitem{CZ90}V.L.Chernyak and I.R.Zhitnitsky, \npb{345}{1990}{137}.
%
\bibitem{EfrRad80}A.V.Efremov and A.V.Radyushkin, \plb{94}{1980}{245};\\
 S.J.Brodsky and G.P.Lepage in:
{\em Perturbative Quantum Chromodynamics}, Ed. by A.H.Mueller,
 World Scientific, 1989.
%
\bibitem{CZ84}V.L.Chernyak and A.R.Zhitnitsky, \prep{112}{1984}{173}.
%
\bibitem{BBr96}P.Ball and V.M.Braun, \prd{54}{1996}{2182}.
%
\bibitem{BBrKT}P.Ball, et al., \npb{529}{1998}{323};\\
               P.Ball and V.M.Braun, \npb{543}{1999}{201}.
%
\bibitem{BBr98}P.Ball and V.M.Braun, \prd{58}{1998}{094016}.
%
\bibitem{BM98}A.P.Bakulev and S.V.Mikhailov, \plb{436}{1998}{351}.
%
\bibitem{Rad98}A.V.Radyushkin,  Preprint \hepph{9811225}.
%
\bibitem{BJ97} V.M.Belyaev and M.B.Johnson, \prd{56}{1997}{1481}.
%
\bibitem{PPRGW}V.Yu.Petrov et al., \prd{59}{1999}{11418}.
%
\bibitem{CLEOggpi}CLEO Collab., J. Gronberg et al., \prd{57}{1998}{33}.
%
\bibitem{Eletsky}T.M.Aliev and V.L.Eletsky, \sjnp{38}{1983}{936}.
%
\bibitem{BBBDosch}E.Bagan et al., \plb{278}{1992}{457}.
%
\bibitem{PDG} PDG Collab., \epjc{3}{1998}{1}.

\end{thebibliography}
\end{document}